\documentclass{aastex6}
\begin{document}

\def\apj{{ApJ}}
\def\mnras{{MNRAS}}
\def\aa{{A\&A}}
\def\Nature{{Nature}}
\def\GCN{{GCN Circ}}
\def\PRD{{Phys. Rev. D}}
\def\PRL{{Phys. Rev. Lett}}
\def\etal{{\it et al.~}}



\newcommand{\grb}{GRB~160509A}
\newcommand{\gr}{$\gamma$-ray}
\newcommand{\grs}{$\gamma$-rays}
\newcommand{\gcn}{GCN Circ.}


\shorttitle{Fermi observations of GRB~160509A}
\shortauthors{Tam et al.}

\title{An evolving GeV spectrum from prompt to afterglow: the case of GRB~160509A}

\author{Pak-Hin Thomas Tam\altaffilmark{1}, Xin-Bo He\altaffilmark{1}, Qing-Wen Tang\altaffilmark{2}, Xiang-Yu Wang\altaffilmark{3,4}   }

\affil{$^1$ School of Physics and Astronomy, Sun Yat-sen University, Zhuhai 519082, China\\
$^2$ School of Science, Nanchang University, Nanchang 330031, China \\
$^3$ School of Astronomy and Space Science, Nanjing University, Nanjing 210093, China \\
$^4$ Key laboratory of Modern Astronomy and Astrophysics (Nanjing University), Ministry of Education, Nanjing 210093, China
}
\email{tanbxuan@sysu.edu.cn}

\begin{abstract}
We present the high-energy emission properties of \grb, from its prompt mission to late afterglow phase. \grb~contains two emission episodes: 0--40s and 280--420s after the burst onset ($t_\mathrm{0}$). The relatively high fluence of \grb~allows us to establish an evolving spectrum above 100 MeV. During the first emission episode, the $>$100~MeV spectrum is soft with $\Gamma\geq$~3.0, which can be smoothly connected to keV energies with a Band function with a high-energy cutoff. The $>$100~MeV spectrum rapidly changes to a hard spectrum with $\Gamma\leq$~1.5 after $t_\mathrm{0}+$40s. The existence of very energetic photons, e.g., a 52~GeV that arrives $t_\mathrm{0}+77$~seconds, and a 29~GeV that arrives $t_\mathrm{0}+70$~ks, is hard to reconcile by the synchrotron emission from forward-shock electrons, but likely due to inverse Compton mechanism (e.g., synchrotron self-Compton emission). A soft spectrum ($\Gamma\sim$2) between 300s and 1000s after the burst onset is also found at a significance of about 2 standard deviation, which suggests a different emission mechanism at work for this short period of time. GRB~160509A represents the latest example where inverse Compton emission has to be taken into account in explaining the afterglow GeV emission, which had been suggested long before the launch of Fermi LAT.
\end{abstract}


\section{Introduction}
Since 2008, the Large Area Telescope (LAT) aboard the Fermi satellite, working at $>$30~MeV energies, has detected over a hundred gamma-ray bursts (GRBs) during the prompt keV-MeV emission phase and/or the afterglow phase. The main characteristics of the $>$100~MeV emission of GRBs before 2011 is described in~\citet{lat_grb_cat}.

The afterglow $>$100~MeV emission is typically characterized by a power law-like decay after a peak time (which sometimes coincides with the prompt emission), and a mean photon index of about 2 for several well studied cases of bright LAT GRBs. The synchrotron radiation of shock-accelerated electrons is usually thought to be the dominant radiation mechanism of the late-time LAT emission up to $\sim$10~GeV~\citep{Zou09,Kumar09}. However, there is a maximum photon energy that synchrotron radiation can reach in the context of Fermi acceleration in the shocks, which in general cannot be much higher than a few GeV in the observer's frame at the deceleration time $t_\mathrm{dec}$~\citep{Piran_2010_external}.

Emission above 10~GeV well after the prompt emission has been detected by the LAT, including GRB~940217~\citep{Hurley_940217}, GRB~130427A~\citep{Fan_130427a}, GRB~130907A~\citep{Qingwen_130907a}, and GRB~131231A~\citep{Liu131231a}. For the very bright and very long $>$100~MeV afterglow of GRB~130427A, inverse Compton radiation was argued to be responsible for the very energetic photons seen especially at late times\citep{Tam130427a,latteam_grb130427a,Liu130427a}, again mainly based on the above maximum synchrotron photon energy argument. The $>$100~MeV emission from GRB~131231A is also well described by a hard power-law with the photon index ($\Gamma\approx$1.5) in the first $\sim$1300 s after the trigger and the most energetic photon has an energy of about 62 GeV, arriving at $\sim$520~s post-trigger~\citep{Liu131231a}.

The relatively small collection area of the LAT has limited the study of such energetic photons (e.g., $>$a few GeV) to the relatively bright GRBs. We note a recent work by~\citet{sample_latgrbs} who investigates the radiation mechanisms of the afterglow LAT emission using a large sample of GRBs. In this work, we focus on the very bright \grb, which, similar to GRB~130427A and GRB~131231A, emit several very energetic \grs. 

\section{Properties of \grb}

\grb~triggered several space instruments: Fermi's LAT and GBM~\citep{gcn19403,gcn19411}, MAXI/GSC~\citep{gcn19405}, Konus-Wind~\citep{gcn19417}, CALET~Gamma-Ray Burst Monitor~\citep{gcn19424}, and INTEGRAL/SPI-ACS~\footnote{\url{http://www.isdc.unige.ch/integral/ibas/cgi-bin/ibas\_acs\_web.cgi?month=2016-05}}. In this work, we take the Konus-Wind trigger time as the reference time (i.e., $t_\mathrm{0}=$2016-05-09UT08:58:46.696). The burst consists of a broad, multi-peaked pulse approximately from $t_\mathrm{0}-$10~s to $t_\mathrm{0}+$30~s, followed by several weaker emission episodes until around $t_\mathrm{0}+$380~s~\citep{gcn19417}. In particular, we identified a second emission episode around $t_\mathrm{0}+$280~s to $t_\mathrm{0}+$420~s (see Sect.~\ref{prompt_lc}). The Konus-Wind fluence in the 20 keV to 10 MeV energy band is (2.90$\pm$0.35)$\times$10$^{-4}$~erg~cm$^{-2}$. As seen by the Fermi GBM, the duration of the burst ($t_\mathrm{90}$) is about 371~s~\citep[50-300 keV][]{gcn19411}.

MAXI/GSC was triggered at UT 2016-05-09 09:04:16, i.e., $t_\mathrm{0}+$329.3~s, and measured a photon spectral index of 1.26$\pm$0.16, and the resultant 2--10 keV flux is 2.78$\times10^{-8}$~erg~cm$^{-2}$s$^{-1}$. Based on the non-detection during the next transit at 10:37 UT on 2016 May 9 (around $t_\mathrm{0}+$5893s), the MAXI/GSC team put an upper limit of 20 mCrab, i.e., about 4.8$\times$10$^{-10}$~erg~cm$^{-2}$s$^{-1}$ on the X-ray flux at this time. 

\begin{figure*}
    \epsscale{.7}
   \plotone{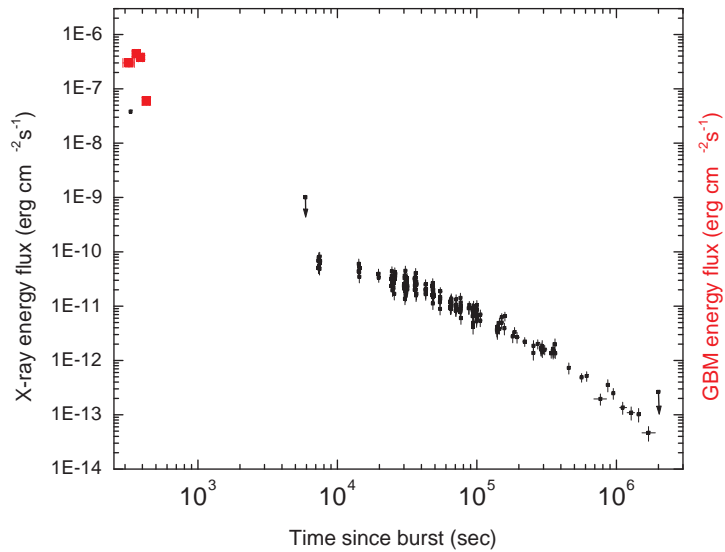}
      \caption{The energy flux from \grb~from several instruments. The red data points corresponds to the 10--1000~keV flux determined from spectral fits using RMFIT from Fermi GBM's NaI (n0) detector. The first two black data points (at 330s and 5900s) were estimated from MAXI/GSC 2--10~keV observations. Latter black data were extracted from the XRT repository for the range 0.3--10~keV.}
         \label{xray_flux}
   \end{figure*}

\emph{Swift}'s X-ray Telescope (XRT) began data-taking of the burst at about $t_\mathrm{0}+7300$s. The XRT Light curve is obtained using the products extracted from the XRT repository~\footnote{http://www.swift.ac.uk/user\_objects/}~\citep{xrt_evan07,xrt_evan09}, and is shown in Fig.~\ref{xray_flux} as the data $>$7.2~ks after the burst, together with the energy flux reported for MAXI/GSC and derived from Fermi/GBM at early times.


We also extracted two 0.3--10 keV XRT spectra, corresponding to 7.2~ks to 7.6~ks and 14.2~ks to 76.9~ks after the burst roughly corresponding to the last two time bins of the LAT emission epoch. Both spectra are adequately described by single power laws, with photon index of $\Gamma_\mathrm{X}=1.62^{+0.27}_{-0.25}$ and $\Gamma_\mathrm{X}=1.99\pm0.09$, respectively. Thus, the X-ray spectrum does not evolve significantly, consistent with the analysis of \citet{grb160509a_VLA}.

The optical afterglow was first detected by~\citet{gcn19410} at R.A.=20:47:00.93, Decl.=$+$76:06:29.2 (J2000). This position is used in the analyses presented in this Letter. It is confirmed to be fading by~\citet{gcn19416}. The redshift of the burst was found to be $z\approx1.17$ with the Gemini North telescope~\citep{gcn19419}. At this distance, its isotropic energy release in keV to MeV \grs, $E_{\rm \gamma,iso}$, is about $1.06\times10^{54}$ erg.

At radio wavelengths, VLA~\citep{grb160509a_VLA} has observed \grb~for weeks after the burst and the authors claim evidence of reverse shock emission from these observations.

The HAWC detector observed the GRB over the prompt emission epoch and did not see any significant emission above $\sim$300~GeV~\citep{hawc_160509a}.

\section{The prompt emission}

\subsection{The two main emission episodes}
\label{prompt_lc}
\begin{figure*}
    \epsscale{.95}
   \plottwo{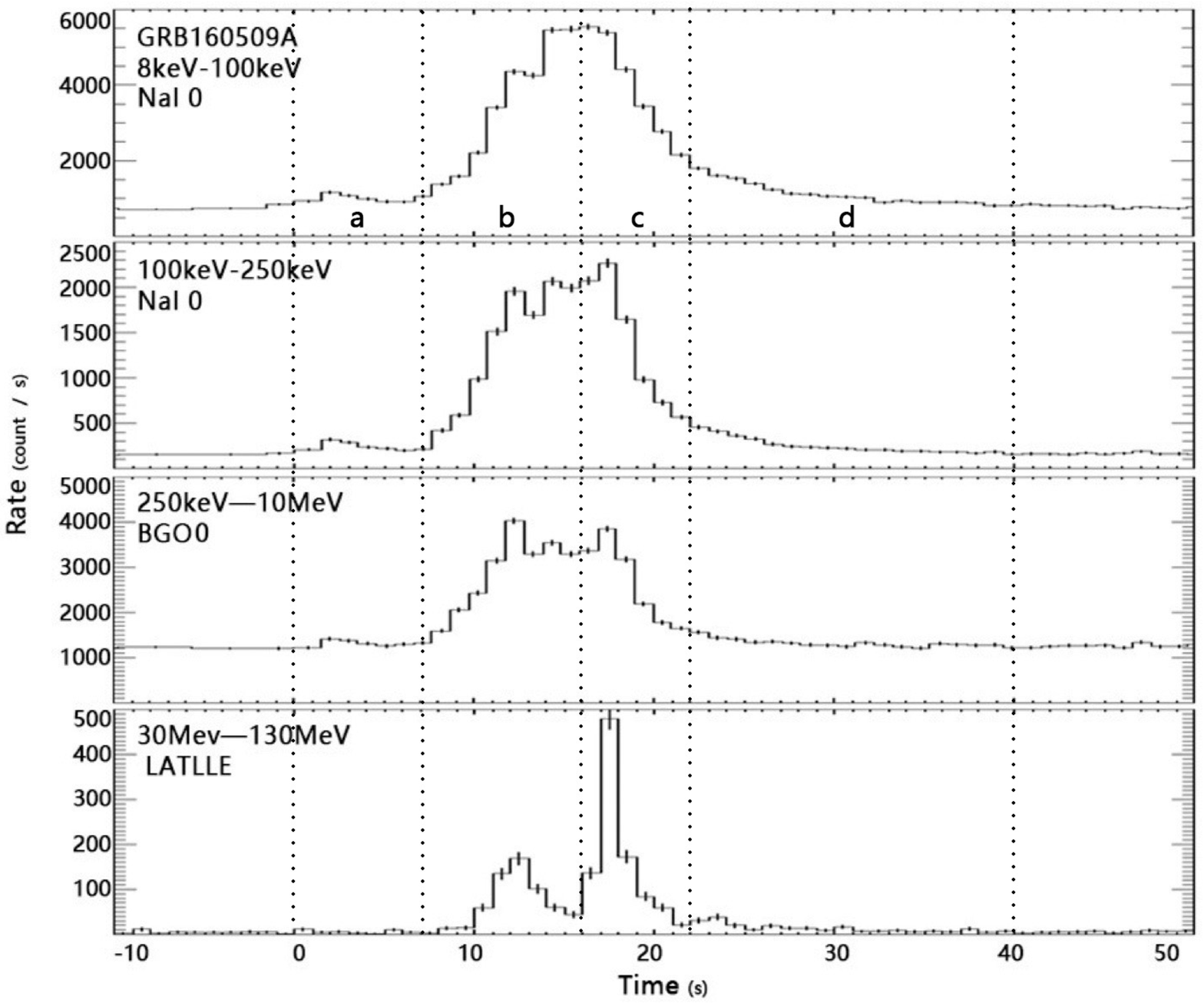}{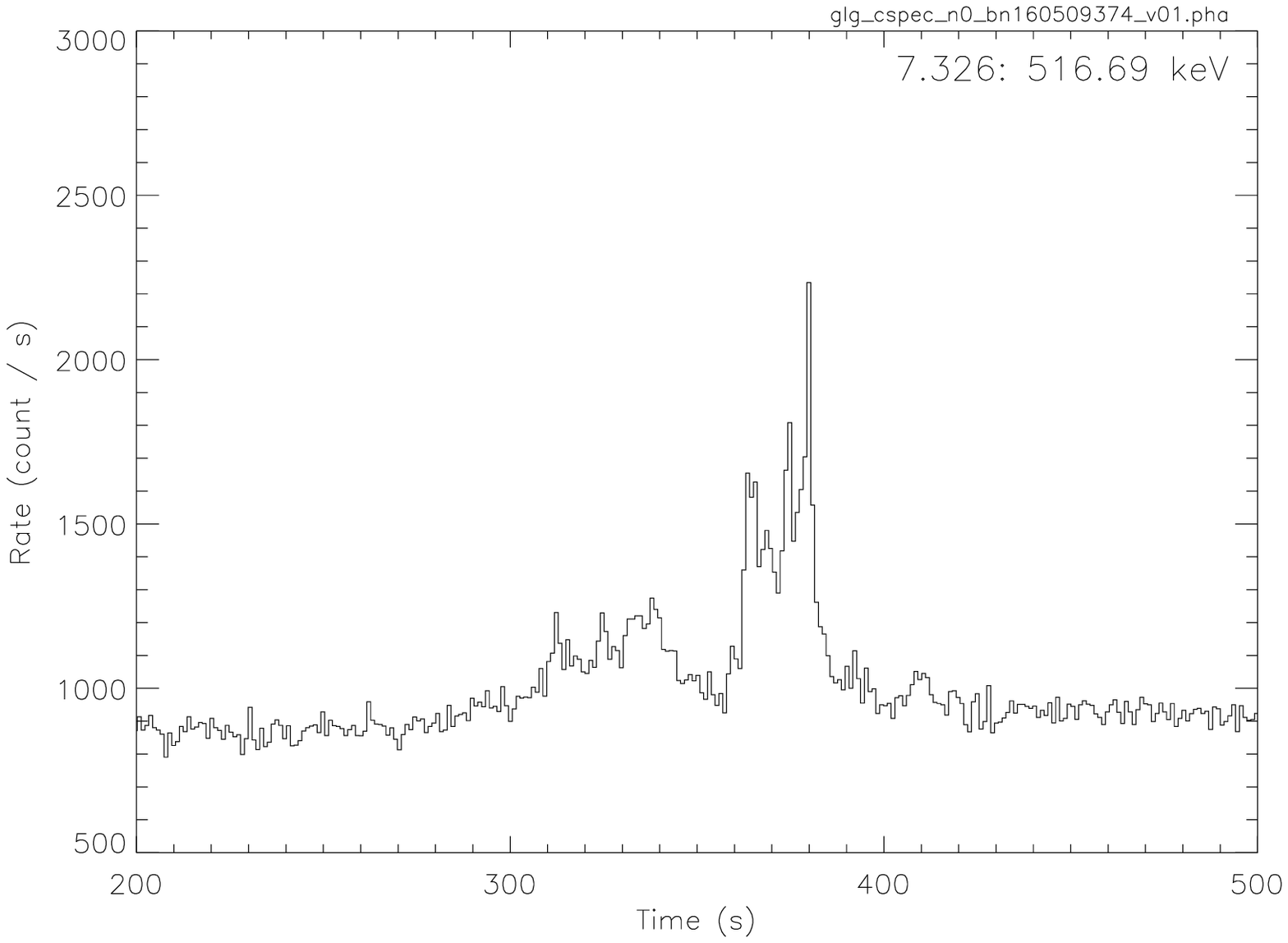}
      \caption{\emph{Left panel:} Energy-dependent light curves for the first emission episode. \emph{Right panel:}  7.3--517 keV light curve for the second emission episode, as seen by the NaI n0 detector.}
         \label{two_sub_bursts}
   \end{figure*}

\grb~is a bright GRB consisting of two emission episodes, 0--40s and 280--420s, separated by a long quiescent period. Fig.~\ref{two_sub_bursts} shows the light curves for each emission episode. For the first emission episode, we can see that LLE has two peaks, the first peak is around $t_0+12$s and the second peak is around $t_0+18$s, while in NaI's n0 and BGO's b0 detectors, the light curve has a rather broad maximum between 12s and 18s. For the second emission episode (from 280s to 420s), the emission was only detected up to around 500~MeV, and thus the BGO and LLE events do not show significant excess during this episode. Hence, the light curve obtained by the NaI's n0 detector is shown.


   \begin{figure*}
    \epsscale{1.}
   \plotone{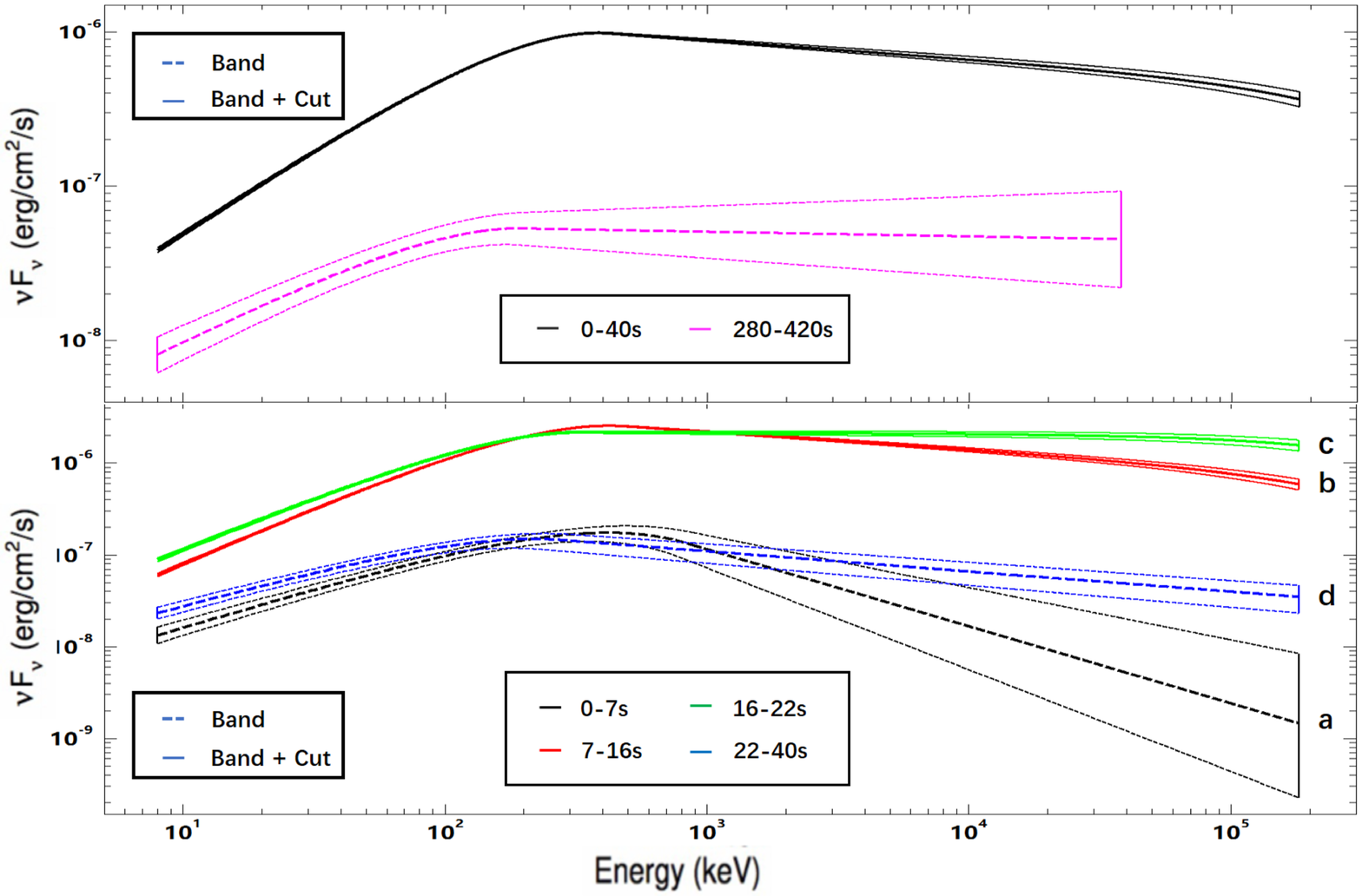}
      \caption{The best-fit model for the first and second emission episodes, as well as different time bins plotted as a $\nu F \nu$ spectrum, derived using data obtained by Fermi's GBM (NaI and BGO detectors) and LAT low energy (LLE) events. The $\pm$1$\sigma$ error contours are propagated from errors on the fit parameters. }
         \label{BandCut}
   \end{figure*}

\subsection{Fermi GBM and LLE spectral analysis}
\label{sect:joint_fit}
During the main burst (or the first emission episode), all GBM detectors saw the emission from \grb. The GRB was also seen in the LAT Low Energy (LLE, 30~MeV--130~MeV) data from 7--28s. To perform spectral fits, we made use of the GBM and LAT data available at the Fermi Science Support Center\footnote{\url{http://fermi.gsfc.nasa.gov/ssc/}}. For 0--40s, we used Time Tagged Event (TTE) data from the good-viewing detectors NaI n0, n3 and the BGO b0 detector, as well as the LLE data. For the second emission episode, we used TTE data from NaI n0 and BGO b0 detectors. The Band function is defined as~\citep{Band93}
\begin{equation}
\label{eqn:band}
f_\mathrm{BAND}(E) = A\left\{
\begin{array}{ll}
	\left(\frac{E}{100\mathrm{ keV}}\right)^\alpha \exp\left[-\frac{(\alpha+2)E}{E_\mathrm{p}}\right]: E<E_\mathrm{c} \, , \\
	\left(\frac{E}{100\mathrm{ keV}}\right)^\beta \exp\left(\beta-\alpha\right) \left(\frac{E_\mathrm{c}}{100\mathrm{ keV}}\right)^{\alpha-\beta}: E\geq E_\mathrm{c} \, ,
\end{array}
\right.
\end{equation}
where 
\begin{equation}
\label{eqn:Ec}
E_\mathrm{c}=\left(\frac{\alpha-\beta}{\alpha+2}\right)E_\mathrm{p} \,
\end{equation}
 and the Band + High Cutoff model is defined as~
\begin{equation}
\label{eqn:bandcut}
f_\mathrm{BAND+Cut}(E) = f_\mathrm{BAND}(E) exp\left(-\frac{E}{E_\mathrm{cut}}\right) .
\end{equation}

In Eqns.~(\ref{eqn:band}), (\ref{eqn:bandcut}) and (\ref{eqn:Ec}), $A$ is the normalization factor at 100~keV in units of ph~s$^{-1}$~cm$^{-2}$~keV$^{-1}$, $\alpha$ is the low-energy power-law photon index, $\beta$ is the high-energy power-law photon index, $E_\mathrm{p}$ is the peak energy in the $\nu F_\nu$ space in units of keV, $E_\mathrm{c}$ is the characteristic energy in units of keV, and $E_\mathrm{cut}$ is the high-energy cutoff in units of keV. Using RMFIT, we found that Band functions satisfactorily describe the first and second emission episodes including smaller time bins indicated in Table~\ref{gbm_spec}, suggesting a similar origin of the GBM and LAT emission for the first emission episode.

Motivated by the LAT analysis at $>$100~MeV during 15--40s which found a soft spectrum of $\Gamma=-3.2\pm0.2$ (see Sect.~\ref{sect:lat_analysis}), we further tested a Band$+$Cut function to fit the different time bins in the first emission episode defined in Table~\ref{gbm_spec}. We found that the Band$+$Cut function can significantly improve the fits for the time bins 0--40s by $\Delta${\it cstat} of 159.4, as well as for the time bins (b): 7--16s and (c): 16--22s. The best-fit model spectra for different time bins during the prompt emission are shown in Fig.~\ref{BandCut}.

The high-energy cutoffs obtained in Table~\ref{gbm_spec} are below 100~MeV, and such spectral cutoffs can be caused by $\gamma$$\gamma$ absorption. For such cutoffs $<$100~MeV, the target photon's energy is comparable to $E_\mathrm{cut}$, i.e., $E_\mathrm{cut}\ga\Gamma_\mathrm{b}^2m_\mathrm{e}^2c^4/[E_\mathrm{cut}\left(1+z\right)^2]$, and the bulk Lorentz factor $\Gamma_\mathrm{b}$ can be estimated by~\citep{Li 2010}
\begin{equation}
\label{eqn:Lorentz factor}
\Gamma_\mathrm{b} \approx \frac{E_\mathrm{cut}}{m_\mathrm{e}c^2}\left(1+z\right)  \,.
\end{equation}
The redshift of \grb, $z\approx1.17$, so the bulk Lorentz factors $\Gamma_\mathrm{b}$ can be calculated using this equation for the corresponding time bins, and shown in Table~\ref{gbm_spec}.

\begin{table}
\caption{Model fits of the two emission episodes. \label{gbm_spec}}
\begin{tabular}{l@{}ccccccccccc}
    \tableline
    \tableline
    $t-T_\mathrm{0}\tablenotemark{a}$ & Model & $E_\mathrm{p}$ (keV)& $\alpha$ & $\beta$ & {\it c-stat}/dof & photon flux\tablenotemark{b} & energy flux\tablenotemark{c} & $E_\mathrm{cut}$\tablenotemark{}(MeV) & $\Delta${\it c-stat} & $\Gamma_\mathrm{b}\tablenotemark{d}$\\
    \tableline\tableline 
    \multicolumn{11}{c}{First emission episode} \\
    \tableline\tableline 
0--7 & Band & 422.2$\pm$70& -1.14$\pm$0.05 & -2.84$\pm$0.27 & 414.9/369 & 4.23$\pm$0.11 & 7.12$\pm$0.34 & ... & ... & ...\\ 
7--16 & Band & 461$\pm$8.4 & -0.78$\pm$0.01 & -2.38$\pm$0.01 & 648.9/369 & 39.11$\pm$0.17 & 93.1$\pm$0.5 & ... & ... & ...\\
... & Band+Cut & 425.3$\pm$2.1 & -0.75$\pm$0.01 & -2.19$\pm$0.01 & 562.1/368 & 38.94$\pm$0.35 & 91.6$\pm$0.9 & 56.2$\pm$10.6 & 86.8 & 239$\pm$45\\
16--22 & Band & 355.7$\pm$8.9 & -0.86$\pm$0.01 & -2.15$\pm$0.01 & 572.3/369 & 45.11$\pm$0.22 & 90.8$\pm$0.59 & ... & ... & ...\\
... & Band+Cut & 326.0$\pm$2.3 & -0.83$\pm$0.01 & -2.01$\pm$0.01 & 459.6/368 & 44.95$\pm$0.58 & 90.5$\pm$1.3 & 69.7$\pm$12.2 & 112.7 & 296$\pm$52\\
22--40 & Band & 218.8$\pm$21 & -1.21$\pm$0.03 & -2.22$\pm$0.02 & 484.0/369 & 5.75$\pm$0.07 & 7.35$\pm$0.15 & ... & ... & ...\\
\tableline
0--40 & Band & 410.1$\pm$7.0 & -0.89$\pm$0.01 & -2.27$\pm$0.01 & 768.5/369 & 18.79$\pm$0.06 & 38.7$\pm$0.2 & ... & ... & ...\\
... & Band+Cut & 384.5$\pm$1.8 & -0.88$\pm$0.01 & -2.12$\pm$0.01 & 609.1/368 & 18.75$\pm$0.14 & 38.5$\pm$0.3 & 72.2$\pm$10.6 & 159.4 & 307$\pm$45\\
\tableline \tableline 
 \multicolumn{11}{c}{Second emission episode} \\
 \tableline\tableline 
280--358 & Band & 238.1$\pm$55 & -0.93$\pm$0.10 & -1.92$\pm$0.11 & 463.4/233 & 1.49$\pm$0.04 & 2.53$\pm$0.09 & ... & ... & ...\\
358--420 & Band & 133.8$\pm$22 & -1.23$\pm$0.08 & -2.17$\pm$0.13 & 501.8/233 & 2.92$\pm$0.05 & 3.12$\pm$0.11 & ... & ... & ...\\
\tableline
280--420 & Band & 189.8$\pm$30 & -1.15$\pm$0.06 & -2.03$\pm$0.09 & 661.6/233 & 2.13$\pm$0.03 & 2.80$\pm$0.07 & ... & ... & ...\\
\tableline
\end{tabular}
\tablenotetext{a}{Time interval; in units of s} 
\tablenotetext{b}{10--1000~keV; in units of photons/(s~cm$^2$)} 
\tablenotetext{c}{10--1000~keV; in units of  $\times$10$^{-7}$erg/(s~cm$^2$) }
\tablenotetext{d}{bulk Lorentz factor calculated using Equation~(\ref{eqn:Lorentz factor})}  
\end{table}

\section{Fermi LAT data analysis and results}
\label{sect:lat_analysis}
The angle of the GRB position is about 32$^\circ$ from the LAT boresight when GBM was triggered and remains within the field of view (FoV) until $\approx t_\mathrm{0}+3000$s. The GeV emission is first seen during the first emission episode, and can be detected as late as about one day after the burst, although the GRB position had been occulted by the Earth several times over the course of a day.

We performed unbinned maximum-likelihood analyzes (\emph{gtlike}) of a 15$\degr$-ROI centered at the GRB position to characterize the spectra of the $>$100~MeV \grs~from the GRB onset to the afterglow phase. 

The Fermi Science Tools v10r0p5 package was used to reduce and analyze the data using standard event selections. We selected photons of energies between 100~MeV and 300~GeV. Using the ``P8R2\_TRANSIENT020\_V6'' events increases the effective collection area, and thus the photon statistics, by $\sim$100\% at 100~MeV, decreasing to $\sim$13\% at 1~GeV, compared to the event class ``P8R2\_SOURCE\_V6''~\footnote{see, e.g., \url{http://www.slac.stanford.edu/exp/glast/groups/canda/lat\_Performance.htm}}. So, we selected this event class for time bins lasting less than 300s (i.e., all time bins between $t_\mathrm{0}$ and $t_\mathrm{0}+$400~s). For longer bins, the background becomes higher and we selected the events classified as ``P8R2\_SOURCE\_V6''. The instrument response functions for the corresponding event classes were used. To reduce the contamination from Earth albedo $\gamma$-rays, we excluded events with zenith angles greater than 100$^\circ$.

The $>$100~MeV photon spectrum from \grb~is assumed to be a single power law, defined as:
\begin{equation}
\frac{dN}{dE} = N_0 \left(\frac{E}{E_0}\right)^{-\Gamma}.
\end{equation}
The Galactic (gll\_iem\_v06.fits) and the isotropic components (iso\_P8R2\_SOURCE\_V6\_v06.txt), as well as sources in the third Fermi catalog~\citep{lat_3rd_cat} were included in the background model for time bins after $t_\mathrm{0}+$400~s. The model includes 3FGL sources out to 15 degrees, while the spectral parameters of sources with detection significance below 10, or variability index below 70, are fixed. Essentially, only the normalization factors of 3FGL~J2005.2$+$7752, 3FGL~J2010.3$+$7228, and the two diffuse components are allowed to vary. For time bins before $T_\mathrm{0}+400$s, the isotropic component suffices to describe the background photons, due to the dominance of the GRB emission over other sources in the ROI during these short-duration intervals. For the first three bins, the normalization factor of this isotropic component is fixed to unity. The photon index of the first data point (0--7s) was fixed at 3.5 to derive the 90\% confidence level upper limit.

The derived light curve and the evolution of the photon index ($\Gamma$) for the 0.1--100~GeV emission up to one day after \grb~is shown in Figs.~4(a) and 4(b).

\begin{figure*}
    \epsscale{1.1}
   \plottwo{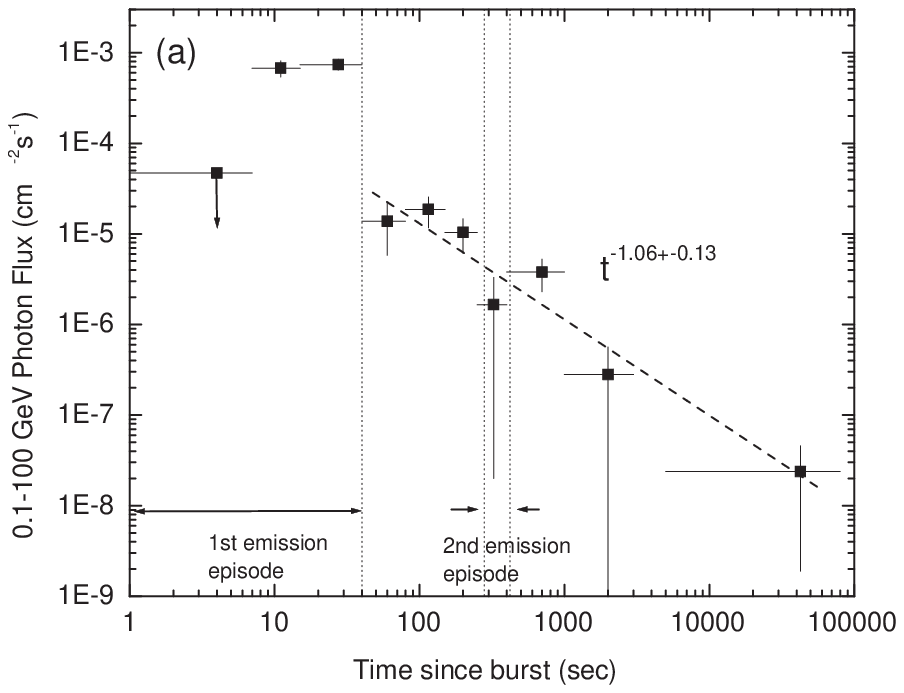}{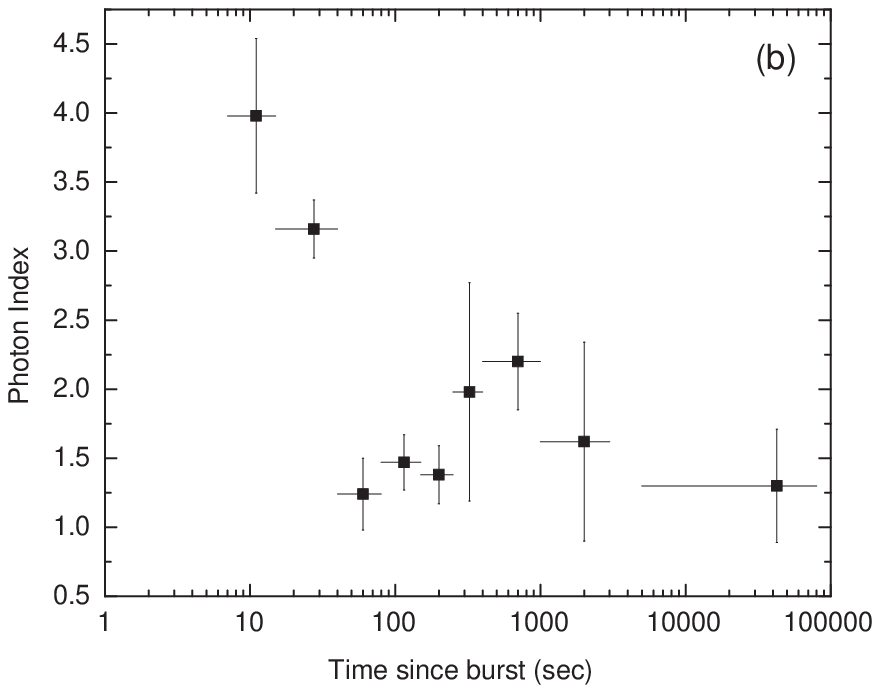}
   \plottwo{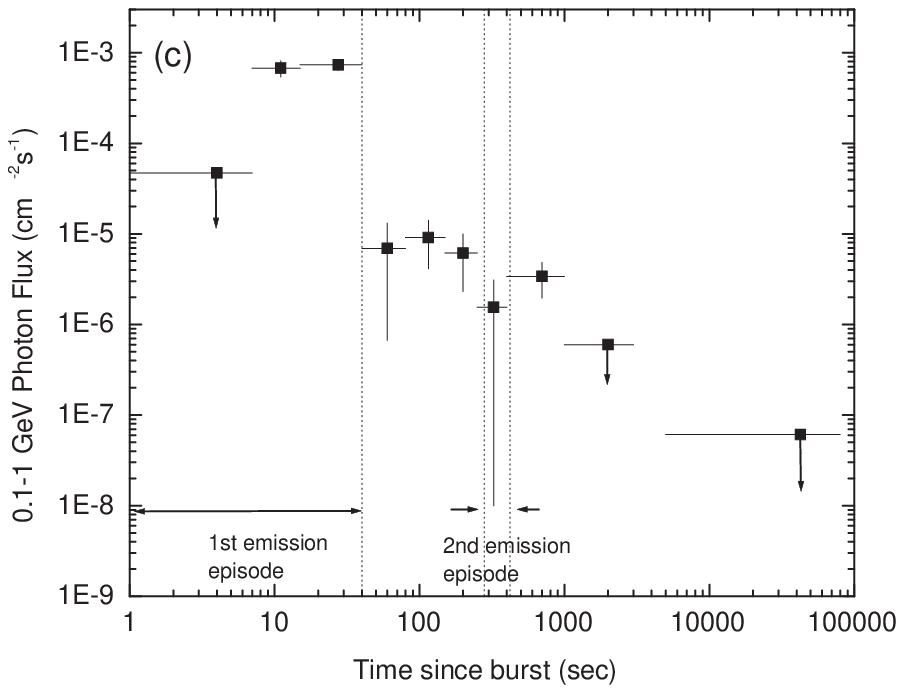}{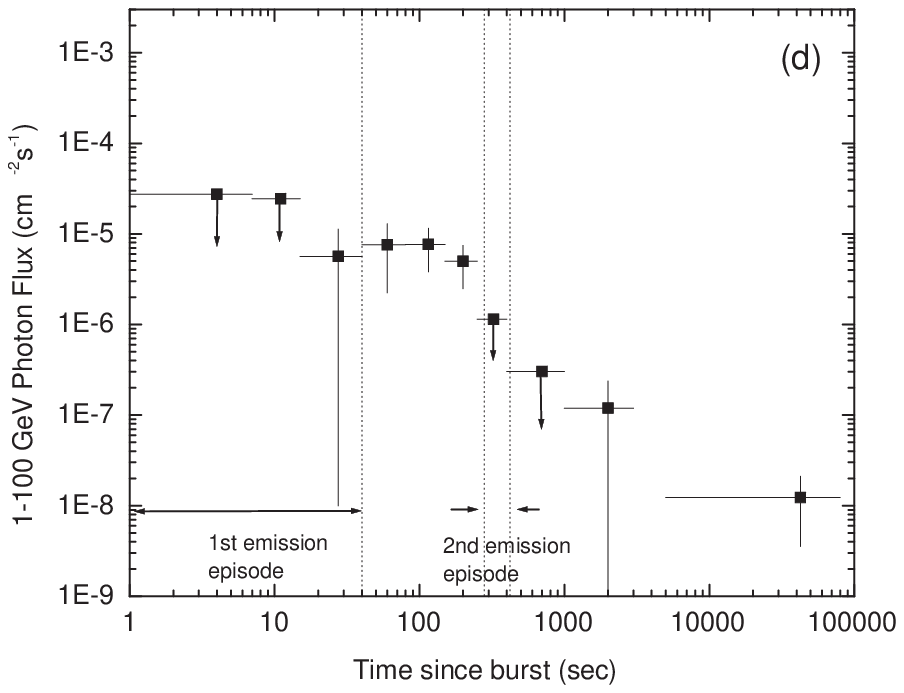}
      \caption{\emph{(a)} The photon flux in the 100~MeV to 100~GeV band, derived from the LAT observations of \grb. The dashed line indicates the power-law fit $F_\nu\propto\,t^{-1.06\pm0.13}$ for LAT temporal decay from 40s to 80ks. The first and second emission episodes are also indicated. \emph{(b)} Evolution of the photon index of the 0.1--100~GeV emission. \emph{(a)} The photon flux in the 100~MeV to 1~GeV band. \emph{(d)} The photon flux in the 1~GeV to 100~GeV band. When there is no detection, the 90\% confidence-level upper limits are shown.}
         \label{lat_flux}
   \end{figure*}

We summarise the LAT emission properties at different times:

\begin{enumerate}
\item {\it 0--40s}: the joint GBM/LLE analysis presented in Sect.~\ref{sect:joint_fit} suggests a similar origin of the GBM and LAT emission. We note that the LAT photon index above 100 MeV is $\Gamma=-3.2\pm0.2$ for the time bin 15s--40s, which is significantly softer than $\beta$ of $\approx$2.2 obtained by the above joint GBM/LLE analysis for similar time bins. This is consistent with the better fit by a Band$+$Cut function (c.f. Sect.~\ref{sect:joint_fit}). The cutoff during this first emission episode has been briefly mentioned by~\citet{Kocevski16}.
\item {\it 40--300s}: After the first emission episode, the LAT emission quickly changes to a hard spectrum ($\Gamma=1.42\pm0.12$ for the period 40s--300s\footnote{we also tested the robustness of the hard spectrum by also allowing the isotropic emission component to vary or keep it fixed, and obtained $\Gamma=1.42\pm0.12_\mathrm{stat}\pm0.06_\mathrm{sys}$}). We note that the highest energy photon (52~GeV) detected from \grb~comes during this period (77s after burst onset).
\item {\it 300--1000s}: There is a modified LAT emission between 300s--1000s (as compared to the hard spectrum seen before and after). This is identified via the soft spectrum ($\Gamma=2.2\pm0.3$) and the mini-bump in the LAT light curve (c.f. Fig.~\ref{lat_flux}). To estimate the significance of the soft spectrum between 300s--1000s, as compared to the spectra seen before and after, we compare the index with the one obtained for the time 40s--80000s, which is ($\Gamma=1.5\pm0.2$), and the difference is about 2 standard deviation.
\item {\it after 1000s}: The photon index for this epoch is $\Gamma=1.4\pm0.3$, which is again very hard. Noticeably, no emission below 1~GeV is seen, c.f. Fig.~\ref{lat_flux}(c).
\end{enumerate}

Because of the spectral evolution, we also plot the light curves for two energy bands: 0.1--1~GeV and 1--100~GeV, as Figs.~4(c) and 4(d).
It can be seen that during the 300--1000s time bin, the emission is seen only below 1~GeV and not above 1~GeV. In contrast, the emission after 1000s is dominated by $>$1~GeV photons.

\section{Discussion}

\grb~contains two emission episodes during the prompt phase, as well as a temporally extended GeV emission. The major finding of our current work is the changing spectrum of \grb~above 100 MeV, from the prompt emission (i.e., within $t_\mathrm{90}$), to the afterglow emission (i.e., up to a day after the burst, see Fig.~\ref{lat_flux}). 

\subsection{On the origin of hard GeV afterglow}
We established a hard spectrum ($\Gamma\leq$~1.5) which is seen during 40--300s after the burst onset and after 1000s. The hard spectrum, together with the 29~GeV photon arriving at 70~ks after burst, is difficult to reconcile by the synchrotron radiation of the forward shock electrons which is usually used to explain GeV emission with $\Gamma\leq$~2~\citep[e.g.,][]{Zou09,Kumar09}. Inverse Compton (IC) emission can play a significant role here~\citep[as suggested years ago by, e.g.,][]{Sari01,Zou09}.

The LAT emission in both epochs can be due to synchrotron self-Compton (SSC) emission of forward-shock electrons, as is suggested to explain the $\Gamma\sim$1.5 LAT spectrum for GRB~130427A and GRB~131231A~\citep[e.g.,][]{Liu130427a,Liu131231a}. \citet{sample_latgrbs} has presented light curves and spectra of 24 afterglows seen by the LAT and identified hard spectra above certain energies, i.e., 0.1--3~GeV for the GeV afterglow from a number of GRBs. For the observed hard spectrum $\Gamma\leq$~1.5 of \grb~and assuming the electron spectral index, $p$, to be $\sim$2--3, the electrons should be slow-cooling (i.e., $\nu_\mathrm{m,ssc}<\nu<\nu_\mathrm{c,ssc}$) and not fast-cooling (i.e., $\nu_\mathrm{c,ssc}<\nu<\nu_\mathrm{m,ssc}$), since fast-cooling electrons would produce a soft spectrum. For slow-cooling electrons, we have $F_\nu\propto\,t^{(11-9p)/8}$ in the ISM case and $F_\nu\propto\,t^{-p}$ in the wind case. Putting p$\sim$2.1, the ISM model is in agreement with the power-law decay of the LAT flux $F_\nu\propto\,t^{-1.06}$ and the spectrum $\nu^0.5$ via $F_\nu\propto\,\nu^{-(p-1)/2}$. 

\subsection{On the origin of the possible short-duration soft GeV emission}
We also found evidence of a modified LAT emission between 300s-1000s (as compared to the hard spectrum identified above). This modified spectrum is mainly manifested via the soft spectrum ($\Gamma\sim$2) and, to a lesser extent, by the mini-bump in the LAT light curve between 300--1000s. Given the lack of simultaneous multi-wavelength observations, we only speculate on the possible origin of this soft emission.

The simplest explanation can be synchrotron emission from the external shock electrons. A major issue with these is why it only dominates at a relatively short time period (300--1000s) but not the whole GeV afterglow epoch (most of which is rather dominated by a hard spectrum). Though there is a huge gap of X-ray observations between the MAXI data at 330s and XRT data at 7.2~ks, it is probably conceivable that the X-rays can be a combination of fast-decay, shallow decay/plataeu, and/or X-ray flares between 330s and 7.2~ks. Indeed, the second emission episode is seen during 280--420s, so the central engine activity can last at least until $\sim$420s.

This extra soft component can also be a result of external inverse-Compton (EIC) processes~\citep{Wang06,Galli07}. In general, the peak energy of the IC emission and the seed photons are related by $\epsilon_\mathrm{p,IC}\sim\, 2\gamma^2_\mathrm{m}\epsilon_\mathrm{seed}$, where $\gamma_\mathrm{m}$ is the characteristic Lorentz factor of the forward-shock electrons, which (in the comoving frame) is given by $\gamma_\mathrm{m}=1.8\times10^3(p-2)/(p-1)\epsilon_\mathrm{e}\,(\Gamma_\mathrm{sh}-1)$~\citep{Wang06}, where $\epsilon_\mathrm{e}$ is the equipartition factor of electrons and $\Gamma_\mathrm{sh}$ is the Lorentz factor of the shock that accelerates the electrons, which can be about 10 to 100 several hundred seconds after the burst. Assuming typical shock parameters, \citet{Fan08} predicted a {\it delayed} sub-GeV component caused by an UV/X-ray flare having seed photon energy of 0.2~keV. While the peak time and duration of this observed soft component is roughly consistent with the EIC emission, a hypothetical UV/X-ray flare had to happen around 300s after the burst onset. One may, however, speculate on an extrapolation of the second emission episode (t$\sim$280--420s) seen by the Fermi/GBM down to UV/X-ray energies. A major drawback of the EIC scenario is that there is no optical or X-ray observations at this time period; it's not possible to relate this additional component to any simultaneous X-ray activities. We do note that late central engine activities (e.g., X-ray flares) could have happened around this time, similar to many cases in other GRBs with early XRT observations~\citep[e.g.,][]{xray_flare}. Indeed, X-ray flares were found to be temporally coincident with the LAT emission of GRB~100728A~\citep{He2012}.

The redshift of \grb, $z\approx1.17$, put it at a distance whose very high energy emission could have been detected~\citep{Xue09}. Indeed, the detection of a 52~GeV photon (which arrived 77 seconds after the GBM trigger) and a 29~GeV photon  (which arrived 70 ks after the burst) are consistent with most models of the extra-galactic background light (EBL) at this redshift~\citep[e.g., see Fig.~1 of][and references therein]{pass8_grbs}. This also verifies that EBL correction on the multi-GeV spectrum for \grb~is not important up to the highest energies we analyzed here.

Having higher sensitivities at the low energy threshold of several tens of GeV, the up-coming Cherenkov Telescope array (CTA) and LHAASO, may be able to detect photons in the 10--100~GeV energy band during prompt and/or afterglow phases of a GRB. Simultaneous low-energy (e.g., X-rays, UV, optical) coverage is also crucial to discriminate the emission mechanism of these energetic photons.

\acknowledgments
This research made use of data supplied by the High Energy Astrophysics Science Archive Research Center (HEASARC) at NASA's Goddard Space Flight Center, and the UK Swift Science Data Centre at the University of Leicester. PHT is supported by the National Science Foundation of China (NSFC) grants 11633007 and 11661161010. TQW is supported by the NSFC under grants 11547029 and the Youth Foundation of Jiangxi Province (20161BAB211007).

\end{document}